\documentclass[sigconf]{acmart}

\usepackage{wrapfig}
\usepackage{tabularray}
\AtBeginDocument{%
  \providecommand\BibTeX{{%
    \normalfont B\kern-0.5em{\scshape i\kern-0.25em b}\kern-0.8em\TeX}}}

\copyrightyear{2024} 
\acmYear{2024} 
\setcopyright{acmlicensed}
\acmConference[CHI '24]{Proceedings of the 2024 CHI Conference on Human Factors in Computing Systems}{May 11--16,
  2024}{Honolulu, Hawai'i}
\acmBooktitle{Proceedings of the 2024 CHI Conference on Human Factors in Computing Systems (CHI '24)}
\acmPrice{15.00}
\acmDOI{10.1145/3613904.3642451}
\acmISBN{00}
\usepackage{acmart-taps}
\begin{document}

\title[Algorithms and Power in Canadian Higher Education]{"This is not a data problem": Algorithms and Power in Public Higher Education in Canada}

\author{Kelly McConvey}
\orcid{0000-0002-1320-7401}
\affiliation{%
    \institution{University of Toronto}
    \streetaddress{27 King's College Circle}
  \city{Toronto}
  \state{Ontario}
  \country{Canada}
  \postcode{43017-6221}}
\email{kelly.mcconvey@mail.utoronto.ca}

\author{Shion Guha}
\orcid{0000-0003-0073-2378}
\affiliation{%
  \institution{University of Toronto}
  \city{Toronto}
  \state{Ontario}
  \country{Canada}}
\email{shion.guha@utoronto.ca}

\renewcommand{\shortauthors}{McConvey and Guha}

\begin{abstract}
Algorithmic decision-making is increasingly being adopted across public higher education. The expansion of data-driven practices by post-secondary institutions has occurred in parallel with the adoption of New Public Management approaches by neoliberal administrations. In this study, we conduct a qualitative analysis of an in-depth ethnographic case study of data and algorithms in use at a public college in Ontario, Canada. We identify the data, algorithms, and outcomes in use at the college. We assess how the college's processes and relationships support those outcomes and the different stakeholders' perceptions of the college's data-driven systems. In addition, we find that the growing reliance on algorithmic decisions leads to increased student surveillance, exacerbation of existing inequities, and the automation of the faculty-student relationship. Finally, we identify a cycle of increased institutional power perpetuated by algorithmic decision-making, and driven by a push towards financial sustainability.

\end{abstract}

\ccsdesc[500]{Human-centered computing~Empirical studies in HCI}

\keywords{Human-Centered Machine Learning, Artificial Intelligence, Higher Education}

\received{September 2023}
\received[revised]{December 2023}
\received[accepted]{January 2024}

\maketitle
\section{Introduction}
Over the past decade, advancements in technology have led to a substantial increase in the availability of data related to students, their behaviors, and their environments. Higher Education Institutions (HEIs) can now harness more data, use it in targeted and real-time ways, and leverage it to enhance organizational processes and outcomes \cite{souto-otero_governing_2016}. The expansion of data collection and data-driven practices by post-secondary institutions has occurred in parallel with the adoption of New Public Management approaches and Knowledge Capitalism \cite{spraakman_new_2009, ramlal_new_2021} policy choices made by governments.

As Higher Education Institutions (HEIs) adopt neoliberal governance strategies, there has been an institutional push towards a greater emphasis on metrics, accountability, and key performance indicators (KPIs) \cite{olssen_neoliberalism_2005}. This has led to the growing use of educational data mining, learning analytics (LA, data collected about a student's activity in the Learning Management System)  \cite{abdul_jalil_learning_2021}, and the development of algorithms for decision-making, predictions, and personalization \cite{bajpai_big_2017}. Along with this trend, both the potential for algorithmic harm and expanding use of student data within HEIs has also risen \cite{marcinkowski_implications_2020, williamson_learning_2021}.
The use of student data for data-driven and algorithmic decision-making is associated with a number of ethical challenges, such as student surveillance and privacy, fairness and equity, and interpretation of data \cite{slade_learning_2013}. Despite the concerns around these practices, there is limited research on learning analytics and algorithmic decision-making as socio-technical systems within HEIs. McConvey et al. \cite{mcconvey_human-centered_2023} conducted a comprehensive review of the existing literature on the design and development of algorithms for use in higher education. The authors found a critical lack of human-centeredness in their design practices and identified a need for more empirical research on how these algorithms are used in practice. Human-centered algorithm design has been frequently proposed in SIGCHI research as a means to incorporating human and social interpretations into the design of algorithmic systems \cite{kim_human-centered_2021,razi_human-centered_2021, scheuerman_human_2023, aragon_human-centered_2022, saxena_human-centered_2020, shneiderman_human-centered_2022}. Additionally, while there has been much research proposing the use of algorithms in HEIs, there is a considerable gap in understanding how these algorithms function within an institution and how the algorithms align with broader institutional goals. Consequently, we ask the following research questions in this paper: 
\begin{itemize}\item \textbf{(RQ1)} Why are student-data-driven and algorithmic decision-making used within higher education? 
\item \textbf{(RQ2)} How are data and algorithms currently designed, used and perceived by the various stakeholders of the Higher Education Institution (HEI)? \end{itemize}
To address these questions, we conducted an ethnographic case study of a public college in Ontario, Canada. We interviewed 33 stakeholders, from students to VPs of various areas, observed a data learning-circle meeting, and reviewed datasets, reports and systems shared with us. To answer \textbf{(RQ1)}, we identified the student data sources, student data-driven systems and algorithms trained on student data in use at the college, and the outcomes for which they are used. For \textbf{(RQ2)}, inductively coded and analysed the transcripts and documents to establish how the college's processes and relationships support those outcomes and highlight the different stakeholders' perceptions of the college's data-driven systems. We then used the ADMAPS framework \cite{saxena_framework_2021} to assess whether and how the college's practices aligned with the identified ADMAPS dimensions of human discretion, bureaucratic processes, and algorithmic decision-making. 
We found that data-driven and algorithmic decision-making are adopted in response to internal and external (governmental and policy) pressures towards financial sustainability: (a) resource allocation as a way of managing scarcity, (b) revenue forecasting and program development through enrollment prediction, and (c) potentially as a source of revenue through commercialization of algorithms \textbf{(RQ1)}. While algorithms are currently limited in use, the college plans to expand its machine learning capacity, requiring new sources of student data. We found that these factors impact students through increased tracking and monitoring of learning activities, exacerbation of existing inequities, and automation of the faculty / student relationship \textbf{(RQ2)}. Together, they work to replicate the college's existing power structures, enabling the college to further their financial sustainability measures and repeat the pattern.

This paper makes two unique research contributions related to the use of data and algorithms in higher education. Drawing on the field notes of our ethnographic case study, we provide insight into the data-driven and algorithmic decision-making practices of a public college. In the following section, we highlight the outcomes the college hopes to achieve through its use of data and algorithms, as well as the data and algorithms themselves. We go on to describe the relationships of the college's divisions and data-stakeholders, its data processes, and stakeholder perceptions of its data systems. Secondly, we identify \textit{
The ASP-HEI Cycle - Algorithms, Student Data, and Power in Higher Education Institutions}, and illuminate how the prioritization of financial sustainability drives the adoption of data-driven and algorithmic practices that ultimately maintain and expand existing power structures within HEIs. 

\section{Related Work}
\subsection{Public Interest Technology at SIGCHI}
Algorithms are increasingly being viewed as an alternative approach to the decision-making traditionally made by street-level bureaucrats \cite{alkhatib_street-level_2019, ammitzboll_flugge_street-level_2021, clancy_reconciling_2022}. These "street-level algorithms" have raised concerns around accountability, transparency, and equity, as tasks traditionally held by street-level bureaucrats are automated \cite{ammitzboll_flugge_street-level_2021}. As stated by Ammitzbøll Flügge et al., "human cases may have characteristics or novelty that cannot be encoded" \cite{flugge_algorithmic_2020}. There exists a growing body of research in the SIGCHI community concerned with the social impact of algorithmic decision-making in high-stakes domains \cite{chancellor_who_2019, delgado_uncommon_2022, flugge_algorithmic_2020, kim_human-centered_2021, razi_human-centered_2021, saxena_human-centered_2020} and the increasing use of AI-based decision support (ADS) systems for organizational decision-making in the public sector \cite{levy_algorithms_2021}. Within this work, researchers have raised significant concerns about the potential for these algorithmic systems to exacerbate existing social issues \cite{kuo_understanding_2023}. Recent research in HCI has explored, for example, the impact of outcome measurement error as raised by Guerdan et al. \cite{guerdan_counterfactual_2023}, issues of data quality in machine learning by Jarrahi et al. \cite{jarrahi_principles_2023}, how definitions of risk are formalized within algorithmic decision-making systems by Saxena et al. \cite{saxena_rethinking_2023}, and issues of validity surrounding the use of predictive tools in complex, real-world decision-making \cite{coston_validity_2023}. 

Concerns have also been raised more broadly around the nature of risk prediction algorithms as fundamentally deficit-based and in turn, likely to move decision-making away from positive outcomes \cite{saxena_framework_2021, holten_moller_shifting_2020}. This concern is particularly salient within the domain of public services where algorithmic decision-making is focused on improving the condition of an individual but balanced with the organizational goals of efficiency and cost savings \cite{holten_moller_shifting_2020}. Alkhatib and Bernstein highlight issues of algorithmic decision-making replacing street-level bureaucrats: when dealing with novel cases, bureaucrats are able to "refine their understanding of the policy" but algorithms in the same scenario will "execute their pre–trained classification boundary, potentially with erroneously high confidence" \cite{alkhatib_street-level_2019}. These issues and tensions are apparent in the algorithms used in public higher education \cite{brown_whose_2020}, as well. A recent review at SIGCHI finds that algorithms are being increasingly adopted by HEIs, but despite recently raised concerns by the research community, little is being done to mitigate the risks and potential harms of these models \cite{mcconvey_human-centered_2023}.

\subsection{Algorithms and Power}
Existing literature on digital surveillance provides important insight into the relationship between data and institutional power. Zuboff highlights anticipatory conformity, an effect of surveillance and measurement where "the pressure of visibility" causes those being surveilled to conform to normative standards of behaviour in order to avoid being singled out \cite{zuboff_age_2020}. According to Lyon, identification could lead to new forms of exclusion, as the surveilled are categorized as potentially undesirable \cite{lyon_surveillance_2010}. Zuboff also warns of instrumentarian power, a new form of power enabled through digital surveillance technologies and the rise of surveillance capitalism, under which, data, control and surveillance work in tandem to uphold capitalism \cite{zuboff_surveillance_2019}.

Despite recent efforts by the SIGCHI research community to center humans in the design of algorithms, discussion surrounding how to reconcile power imbalances among algorithmic stakeholders within algorithm design has been limited \cite{park_power_2022}. While recent work highlights the multitude of issues and concerns in algorithmically-supported decision-making in high-stakes contexts like HEIs, the role of algorithms in maintaining or exacerbating the power dynamics between student and school has not been explored. And just as in the private sector, data, control and surveillance work in tandem to uphold the power of our colleges and universities over students. 

\subsection{Algorithmic Trends in Higher Education}
Prior Education research describes neoliberalism in public institutions as "the dismantling of boundaries between the economic and the social, rendering the latter accessible to the former and making it a source of financial profit" and in doing so, creating demand for the data-driven tools of New Public Management (NPM) \cite{piattoeva_escaping_2020}. 
In HEIs, neoliberalist approaches "have put a new focus on performativity... an emphasis on measured outputs: on strategic planning, performance indicators, quality assurance measures and academic audits" \cite{olssen_neoliberalism_2005}. Key performance indicators and metrics are attractive to administration, carrying with them an "aura of disinterestedness, impersonality, objectivity and universality" \cite{piattoeva_escaping_2020, desrosieres_politics_1998}. These metrics appear transparent and accessible, and therefore in alignment with the values of the institution and democracy itself \cite{piattoeva_escaping_2020}. But numbers reduce, flatten, conceal and strip away context, detail, and the subjective \cite{piattoeva_escaping_2020, prinsloo_black_2020}. In spite of this, data is far from objective as it reflects the interests of "economic, political, social, technological, environmental and legal apparatuses, structures, and elements" \cite{prinsloo_social_2019}.

This is especially concerning to Education researchers as quantification creates a "silent, implicit authority in which the exercise of power is difficult to discern" \cite{piattoeva_escaping_2020}. Further, Piattoeva emphasizes that "quantification, counting, accounting, enumeration and numbering are inherently and simultaneously complex socio-cultural and socio-technical practices. Although often portrayed as robust, objective and neutral, numbers are nevertheless inherently interpretive, fluid and amorphous" and that "how things get measured often influences how they get managed" \cite{piattoeva_escaping_2020}. For Prinsloo \cite{prinsloo_black_2020}, it goes beyond simply an influence on how things are managed, but rather creates a system of governance in HEIs. Prinsloo presents an version of Zuboff's instrumentation power specific to HEIs, stating that "the collection, analyses and use of student data furthermore resembles an ‘algocracy’ described as the normalisation and naturalisation of governance by algorithms that exclude, per se, any disagreement or user interrogation, or other discursive or material options" \cite{prinsloo_black_2020}. 

\section{Methods}
\subsection{Study Overview}
The research site for this paper was Centennial College, one of Ontario's 24 publicly-assisted colleges that grant diplomas, bachelors degrees and certificates for domestic and international students \cite{government_of_ontario_published_2023}. The first author is a full-time employee of the research site.  Publicly-assisted colleges and universities across Ontario have been urged by the provincial government to prioritize financial sustainability, which they define as "“the financial capacity of the public sector to meet its current obligations, to withstand shocks, and to maintain service, debt, and commitment…at reasonable levels relative to both national expectations and likely future income, while maintaining public confidence” \cite{harrison_ensuring_2023}. The province has committed to rewarding financial sustainability and promoting economic growth in HEIs, while continuing its tuition freeze \cite{government_of_ontario_ontario_2023}. The majority of funding for Ontario's colleges comes from student fees, with public funding dropping from 54\% to just 32\% of total funding over the past decade \cite{government_of_canada_trends_2022}. With a Government-issued tuition freeze for domestic students and limited Government support, colleges have turned to international student enrollment to increase revenue with a "six-fold increase in the number of international college students in Ontario" from 2009 to 2019 \cite{government_of_canada_trends_2022}. The province is implementing a performance-based funding model in 2023, initially planned for 2020-2021 but delayed by the global pandemic \cite{government_of_ontario_published_2023}. With it, public funding for three years will be tied to measurements of colleges' annual outcomes.

We conducted a four-month long in-depth ethnographic case study from May to August 2023. Prior to beginning participant interviews, we obtained Research Ethics Board (REB) approval from our research institution and from our research site, we provided participants with a REB-approved consent form and obtained verbal consent. The first author conducted semi-structured interviews with 33 key stakeholders at the college (See Participant Table on Page \aptLtoX[graphic=no,type=html]{4}{\pageref{Participant Table}}), observed a data learning session, and reviewed documents shared throughout the interview process. All but two interviews were conducted remotely with video, one was in person and one was audio only. Interviews were 30-60 minutes in duration. We asked participants a series of questions about the way the college uses data and algorithms, and the surrounding perceptions. 

\subsection{Qualitative Data Analysis}
Interviews were recorded (video where possible) and transcribed verbatim. Field notes and transcripts were processed using the framework described by Emerson, Fritz and Shaw \cite{emerson_writing_2011}, analyzed using open coding by the first author, followed by development of memos, themes and finally, focused codes which were shared among all co-authors. We used open-coding to identify the high-stakes outcomes for which student data-driven and algorithmic decision-making are leveraged and how data and algorithms are currently used within the college's work practices. We then adopted the ADMAPS framework to reveal the complex inter-dependencies between ADMAPS dimensions \cite{saxena_algorithmic_2023}, coding out data to human discretion, bureaucratic processes and algorithmic decision-making \cite{saxena_framework_2021}. Finally, we analyzed the resulting patterns to identify emergent themes to construct a holistic view of the role of data and algorithms in the institution. We chose to use a hybrid approach to coding, beginning with inductive coding before employing the ADMAPS framework deductively. Following standard SIGCHI research practices, we wanted to first understand what themes emerged specific to Higher Education empirically, before steering the quotes towards any particular framework. Then we coded to the ADMAPS framework in order to connect our findings to existing theory in SIGCHI. 
\section{Results}
This section is divided into two main subsections. In \ref{Outcomes, Data, and Algorithms}, we present the college's use of data and algorithms as described by participants. \ref{Outcomes, Data, and Algorithms} is divided further into three subsection: the college's goals and target outcomes for its use of data and algorithms (\ref{Data-Driven Outcomes and Institutional Goals}); the college's collection, ownership, and use of data (\ref{Data Collection, Ownership, and Use}); and the college's algorithmic decision-making practices now and in the future (\ref{Adopting Algorithmic Decision-Making}). These findings outline the \textit{technical specifics} of how data and algorithms are used within the college.

The second subsection (\ref{Relationships, Data Processes, and Institutional Perceptions}) is also divided into three parts. In the first (\ref{Relationships Defined by Data}), we explore participant descriptions of how their relationships within the college are defined by data. Next, we outline the college's processes for acquiring and using data-driven and algorithmic systems (\ref{Data Processes}). Finally, we describe the participants' perceptions of these systems and identify tensions between institutional departments and roles (\ref{Perceptions of Data and Algorithmic Decision-Making}). In this sections, important themes emerge surrounding the \textit{impact} that data and algorithms have across the college. It is these findings that form the basis of our key contribution, the ASP-HEI Cycle detailed in the 
Discussion section.
\begin{table} \label{Participant Table}
\caption{Participants by Role}
\begin{tabular}{ll}
\toprule
\textbf{Role}                  & \textbf{Participants}        \\
\midrule
Academic Chair                 & \textit{P18, P19, P21, P22 }          \\
Advisor                        & \textit{P26   }                       \\
Data Analyst                   & \textit{P25, P30, P31, P39, P41, P47 }\\
Dean                           & \textit{P23, P34, P50    }            \\
Director                       & \textit{P29, P52, P53, P59  }         \\
Learning Technology Specialist & \textit{P45, P46 }                    \\
Manager                        & \textit{P24, P28, P33, P44 }          \\
Professor                      & \textit{P13, P15, P48}                \\
Student                        & \textit{P1, P3, P4, P5   }            \\
Vice President                 & \textit{P35, P37}                 \\
\bottomrule    
\end{tabular}
\end{table}
\subsection{Outcomes, Data, and Algorithms} \label{Outcomes, Data, and Algorithms}
\subsubsection{Data-Driven Outcomes and Institutional Goals} \label{Data-Driven Outcomes and Institutional Goals}

In its 2021-2025 Academic Plan, the college identified data-driven decision-making as an ongoing priority, specifying the need to \textit{"leverage data to inform evidence-based decision-making in how we support student success and retention, graduate employment, quality assurance, and process improvements..."}  \cite{centennial_college_academic_2020}. Growing emphasis on accountability and financial sustainability \cite{government_of_ontario_ontario_2023} have led to the adoption of analytics strategies more often seen in for-profit businesses than the public sector. One Vice President within the college described the need for data-driven practices in higher education as \textit{"...absolutely a must. I spent 20 years in analytics and strategy in the corporate sector, and data is everything. In the higher education sector it's a little bit different, but not as different as as many people think
"} (P35).

Increased data-driven practices and systems in higher education are viewed as imperative and explicitly linked to efficiency, effectiveness, and accountability promised with New Public Management \cite{lane_new_2000}. In practice, the most common outcome for the use of data and algorithmic decision-making, as cited by study participants, was resource allocation. P20, a Dean at the college, describes it as a decade-long trend towards an environment of "\textit{resource scarcity; increasing complexity of student needs}".\label{P20-1} 

As the Academic and Student Experience Divisions face greater challenges, they are met with an expectation of greater efficiency and decreasing resources. Data is viewed as a way to cope with these changes. P35 described the use of data-driven decision making as crucial to managing a more complex environment with fewer resources, \textit{"Part of why data is incredibly important is so that you're using the right resources to do the right work."}\label{P35-1} Data is used within the decision-making processes for the allocation of resources among divisions, staff, schools and even students at an individual level. Careful resource allocation becomes more important when those resources are scarce, as has been increasingly the case in recent years. The push towards financial sustainability, and with it, increased revenue, over the past decades has exacerbated this situation: \textit{"We're still responsible for turning a profit in terms of money to reinvest into further growth, and so utilizing data to maximize our revenue is a critical aspect of what we do... 
    "} (P35). \label{P35-2}

The savings provided from cost-cutting are then reinvested into revenue-generating programs - a financial sustainability cycle with data at its center. While data has been used broadly to drive decision-making across the college, algorithmic decision-making is increasingly being used. Currently, algorithms are deployed for two major outcomes: revenue projections through a series of enrollment prediction algorithms, and resource allocation through a series of algorithms designed to identify at-risk students, as discussed in detail in the section \hyperref[sec:algorithms]{Algorithms}.

\subsubsection{Data Collection, Ownership, and Use} \label{Data Collection, Ownership, and Use}
\label{Data}
The college collects and has access to vast and varied student data, but participants described limited uses for that data currently in practice: \textit{"We have a lot of data. But we don't use a lot of that data yet..."} (P50) 

Participants cited difficulties in knowing what data exists, accessing it in a timely manner, and insufficient capacity for deep analysis as reasons why data isn't used more fully. The result is that deep profiles of students are collected and stored without an explicit purpose, only nebulous plans down the road. 

Issues around who owns the data and who decides how it should be used arise frequently: \textit{"There is rich data. It's which ones do we use? Who's in charge of collecting the data? Who's in charge of distributing it? In charge of future proofing it?"} (P23) As no owner is clearly identified, processes around how it can be used are undefined and ad-hoc. And without clear policy and ownership, no one is responsible for ensuring the data is used in alignment with institutional values and student expectations. 

The same issue exists for the algorithms that are trained on that data and their outcomes, as described by P53, a Director at the college: \textit{"No one seems to know who owns data. I keep asking, hoping someone can give me an answer. 
We don't own the Early Alert System score necessarily, because it's really just made up of student data"} (P53). Adding to the complexity is that differing datasets and metrics are used to measure the same thing across the college. One Dean gave the example of enrollment prediction:

\begin{quote}
    \textit{"We have [the data and research office]'s projections and the International team's projections, so we have two sets of data that sometimes speak a different narrative to us... 
    So, we have three datasets and are trying to figure out which one is the right one at the right time, and that is a bit of insight into Centennial College. Both are valid. 
    "}  (P23) \label{P23-1}
\end{quote} 

Teams are left to determine a single source of truth for their own purposes, which might differ from another division's analysis. This leads to different narratives across the college and misalignment in how best to reach goals. Not only do priorities differ between divisions, but data narratives differ as well. And in an environment of limited resources, the division with the strongest data-driven argument is likely to be prioritized.

A member of the Advising Team describes the problem as one of institutional silos impacting the use of data: \textit{"This is not a data problem. This is like an institutional issue. But the silos are what affects how the data is used..."} (P25)

The college's burgeoning data practices reflect the already existing silos within the institution. And as the college begins to rely more heavily on data and algorithms for decision-making, and meeting and measuring its strategic goals, those institutional issues will be further exacerbated. This increased reliance on student data will, as we see in \ref{surveillance}, lead to increased surveillance. 

\subsubsection{Adopting Algorithmic Decision-Making} \label{Adopting Algorithmic Decision-Making}
\label{sec:algorithms}
In this section, we will describe the models currently in use at the college, with a focus\break on the college's algorithmic Early Alert System. The section is\break divided as follows: 
{Existing Algorithms}, 
{Early Alert System}, 
{Model Building and Architecture}, 
{Model Performance}, 
{Algorithmic Bias},\break and 
{Features}.

\paragraph{Existing Algorithms}\label{Existing Algorithms}
The college is in the early stages of adopting algorithmic decision-making. This move is driven by the data team, with a single data scientist responsible for building and training models for use by other departments. Three algorithmic use-cases exist at the college, each with a series of trained models, with plans to expand use of algorithmic decision-making going forward. A member of the data team described the models in use: 
\textit{"Now it's three [models] but may be more in the future. One is student-level enrollment prediction. The second is program-level enrollment prediction. The third is [to] predict the probability [that] the student can graduate from our college successfully."} (P31)

The two enrollment prediction models are used for institutional planning to predict the number of students that will register in the upcoming academic year. One makes the prediction at the student level - how likely is a given student to enroll in the college? The second predicts the number of students to enroll in each program. This is used for scheduling, staffing, budgeting, and space allocation planning. 

\paragraph{Early Alert System} \label{Early Alert System}
The third model, commonly referred to as the Early-Alert System (EAS), has gone through multiple iterations over many years and was finally implemented in the 2022-2023 academic year. Of the three algorithms in use at the college, it has the most direct impact on students.

\begin{quote}
    \textit{"Each student is given a predictive score that will tell us the degree or the likelihood that they're going to either persist or fail out. We've grouped these students essentially into three different categories: red, yellow, and green. Red, meaning that they're obviously at high risk, requiring the most amount of intervention or the most support. And so this is then filtered to a system called CRM Advise that advisors at the college have access to, and are able to easily identify those students who require intervention..."} (P25) \label{P25-1}
\end{quote}

The college's Success Advisors are assigned a cohort of students each year. Each student in their cohort is grouped into one of the three categories: Red, Yellow or Green. The advisors only see the student's category and cannot view the predicted probability. Separate algorithms were trained for different groups of students, dependent on where they are in their academic career. The Early Alert System scores, and resulting categorization, are updated each semester:

\begin{quote}
    \textit{"We get the predictive model for every semester and it functions in two ways: pre-arrival data and then registered student data. Pre-arrival is for students who have applied to Centennial College, but haven't necessarily registered... And so this is trying to understand,
    what is the chance that they're going to persist if they choose to stay with Centennial College?
    "}  (P25)
\end{quote}

The pre-arrival model predicts the likelihood that an applicant will graduate, should they enroll. The subsequent models are updated each semester with student academic registration and performance data. In many ways, the model is intended to quantify and automate a role that faculty have frequently played for students: identifying and intervening on students who may be struggling \cite{lillis_faculty_2011}.

\paragraph{Model Building and Architecture} \label{Model Architecture}
All of the algorithms used at the college are built using DataRobot, a 3rd party automated machine learning tool (P31). The dataset is loaded into the platform and the target variable is set. The system cleans the data, imputes missing values, and then trains and tests multiple algorithms: \textit{"We use DataRobot to build the machine learning models... the system will recommend the best model. Why we chose [DataRobot] at the beginning, I don't know..."} (P31) \label{P31-2}

DataRobot was initially acquired by a previous data team, and the current team isn't sure why it was selected. P31 describes the development process as hands-off, with little control over how the algorithms are selected and trained, and limited visibility into the process: \textit{"
You don't need to do any coding. And you don't need to deal with missing values, the system will do it for you automatically...The system will recommend the fastest model. That's the system's thing... It's some Blast model. It's like a Tree model. A group of models... It's not directly visible to us, and we don't look at the detail of those algorithms"} (P31).

\paragraph{Model Performance} \label{Model Performance}
The platform evaluates the models' performance but the college shares limited details about model accuracy, as well as what metrics they use to evaluate performance. Decisions around the acceptability of a model's performance are made by the Data Scientist who built the model and in isolation from the model's stakeholders, such as those who will be acting on the model's outcomes or the students impacted by the model's decision. Further, there is no evident policy or guideline for what is an acceptable accuracy level for the current algorithmic use cases. 

According to P31, \textit{"At the beginning, the accuracy is not very high, but we can use this model just to choose which students need more help"} (P31). They indicated that the model is used regardless of the model's questionable performance. And when asked about the specific performance metrics, the participant responded \textit{"Sorry, I can't share [the model accuracy]. I can say it's more than half. If the accuracy is less than 50\%, if the accuracy is less than 60\%, I would say we cannot use this model."} (P31)

Model performance is treated as confidential; information that is controlled by the data team alone. In spite of the college's lack of transparency surrounding model performance, there is a growing perception that increased automation and reliance of algorithmic decision-making will increase the overall accuracy of the data the college uses. One member of the executive team explained,

\begin{quote}
\textit{"You know, because I'm in statistics, I know what a representative sample size is - a number of people wouldn't. So you might do a survey or get a data set and make decisions around it. But the sample size is not statistically significant. So that's an error in judgment. And so by automating a lot of these things, it also increases the accuracy with the data that we're using. 
"} (P35)
\end{quote}

Participants across divisions pointed to accuracy as a metric to support their trust in the model's design, use, and outcomes. For many of the participants we spoke with, automation leads to accuracy and accuracy means it is working as intended. A reliance on accuracy, however, can lead to greater bias as covariate shift occurs and new data no longer resembles training data \cite{chiang_youd_2021}.

\paragraph{Algorithmic Bias} \label{Algorithmic Bias}
While automation is identified as a means to increase accuracy in decision-making, there was little mention of any risks or harms associated with automation, and especially automation through algorithmic decision-making. We identified no formal or ad-hoc processes to identify risks or mitigate harm caused or exacerbated by automation. Additionally, participants expressed a lack of clarity over who owns, and is subsequently responsible for, the decisions of the employed algorithms, should problems arise.

\begin{quote}
    \textit{"I think it's a joint responsibility, really like [the data team] would look at us as the business owners because we are. We commissioned it. We have pushed it. We will be rolling it out. We are responsible for the project and the score and the utilization of it. But they own the the algorithm, so to speak, right? So if there's something wrong with the algorithm, we can't do anything without them. So I guess we own the project, but they own the mathematics that actually have created the score."} (P53) \label{P53-2}
\end{quote}

There are no guidelines or policies for how models and their outcomes must be reviewed to ensure bias is mitigated. The only current method of evaluating the mode bias is automated within the DataRobot platform:

\begin{quote}
    \textit{"Because this is automatic, to change it is a little bit difficult. I cannot. The only thing I can do is to choose different features. I will try several times to see which features are working."} (P31) \label{P31-1}
\end{quote}

The Advising team, who are responsible for using the output of the model to target at-risk students, also lack processes for chequing the bias of the model. When asked how they and the data team work together to ensure that biases or inequalities aren't heightened because of the model, a Director on that team wasn't sure: "I don't [know]. I mean, I trust that [the data team] are doing that." (P53) \label{P53-1}

The absence of defined ownership and processes puts students at risk: without clear guidelines for model performance, bias goes largely unchecked and accuracy is an assumed result of automation free from review or critique.

\paragraph{Features} \label{Features}
The DataRobot platform also automates feature engineering and calculates feature importance measurements. Each version of the EAS is built and trained separately, using variations on the feature set. The pre-arrival model has a more limited feature set, including only what is known about the student at the time they apply. As the student progresses through the college, the model is updated with their academic performance, financial, and registration information, as explained by P25:

\begin{quote}
    \textit{"
    Essentially, the score is not static. It starts off static. But the idea is that the score is going to move each semester for two reasons: one, because we're adding in more data, right? As they complete a course, as they attend more classes, as they pay their tuition. But also, as we start to build out interventions, the goal here is to build these interventions and track them, and to collect the data."}
\end{quote}

The college's approach to building the initial feature set is to include all available data. Anything they know about the applicant or student is used in training. The college plans to expand this data set in the future and potentially include LA data from the college's LMS:

\begin{quote}
    \textit{"All historical [High School data], as well who they are, where they come from, have they graduated. Their status. All demographic information is placed into it. Everything, everything we can find."} (P29)  
\end{quote}

\begin{quote} \label{LMS Data}
    \textit{"Almost everything. And if we can deep mine student data, we could find more data... We haven't used LMS data but that is the next step..."} (P31) \label{P31-3}
\end{quote}

This approach of collecting and including "everything we can find", and plans for the further mining of student data, are in contradiction to approaches from other divisions at the college. P50, a Dean at the college, described the current approach to data collection as problematic:
\begin{quote}
    \textit{"I think a lot of times we get around the table, we're like, "okay, what else should we collect about students?" But then my question is, well with the data that we are collecting, how are we using it? And in terms of asking what else we can collect, why, to what end? For what purpose? And how is that going to support students?"} (P50)
\end{quote}

The purposeful and intentional collection of data is at odds with the current modelling approach of using everything available, beyond the purposes for which it was originally collected. The data team describes their preferred approach to feature selection similarly, as trial-and-error using any data they can access, and allowing the DataRobot platform to determine feature importance: \textit{"No one can tell you if we add new features, if they will be useful. The only thing [we can do] is try..."} (P31).

This approach to feature selection requires a vast and growing amount of student data, as improvements to the model can only be achieved through the addition of new data. This leads to more data collection and with it, increased surveillance of students as we discuss in \ref{surveillance}.

After training the model with demographic and academic data, the [data and research office] team indicated that academic performance was the most predictive feature for those students who are already at the college, as is expected for a model predicting the likelihood of academic success: \textit{"For students [with] some performance at our college, the most important thing is GPA.} (P31)

For the pre-arrival model, where students haven't yet begun their studies at the college, the Advising team described the three combined variables that most contributed to the prediction: 

\begin{quote}
    \textit{"
    One single variable does not contribute to that score. If we put the variables on their own, it looks like they're not very meaningful, but when we combine them so when you look at the season in which they apply, the age and the program, those three together are the most indicative of that score..."} (P25) \label{P25-2}
\end{quote}

These features were a surprise to the advising team who had their own long-used but informal heuristic rubrics for identifying students at risk. The advising team is unclear on the cause of the relationship between student age, season in which they apply and their subsequent success. And as they can't determine the reason from the model output alone, they are unable to develop appropriate interventions.

\begin{quote}
    \textit{"[The data team] said, 'We looked at the data for the last 10 years. And this is what the data is telling us. It's telling us that these are the three factors about students, these are the three criteria that matter most when a student is coming in as to whether or not they're going to be successful.' So it's hard to argue with that.} (P53)
\end{quote}

Here, the model is seen as infallible. The features identified as having the highest predictive power were both inexplicable and opaque for the Advising team responsible for implementing the model's decisions. But despite coming in with extensive experience making decisions similar to those of the model, they are unable to question the model's results. This takes power out of the hands of the student experience team, as discussed in \ref{inequity}.

\subsection{Relationships, Data Processes, and Institutional Perceptions} \label{Relationships, Data Processes, and Institutional Perceptions}
\subsubsection{Relationships Defined by Data} \label{Relationships Defined by Data}
While aligned on the need to move the college forward in its adoption of data-driven practices, participants identified different barriers to progress. A member of the Student Experience team pointed to the centralized nature of the the data team as an obstacle when needing to access data:

\begin{quote}
    \textit{"It's the data office that's responsible for all this at the end of the day. No one in the college, aside from that office, really knows all the data that we have. I have no idea everything that we collect, everything that we have access to and how it's being used, even though I'm considered a data person for the college... 
    "} (P25)
\end{quote}

To overcome these challenges, some departments have begun building data analysis capacity within their own teams. One Dean described the decentralized model adopted within their academic school:

\begin{quote}
    \textit{"
    Because we couldn't get access to some of the data that we needed, we created the dashboard ourselves. So now, by nature of having that, and having some talented people on our team, we are decentralized and it works.
    "} (P23) \label{P23-2}
\end{quote}

Without knowledge of what data are available and the resources for timely analysis, teams across the college are left to either develop their own solutions or make do without. The Business Division, however, identified an unwillingness from the other divisions to accept additional accountability as the cause of their reluctance to adopt greater data-driven processes: \textit{"I would just say that some of the downside is it's a paradigm shift for higher education. You know what the problem is with data is it drives accountability. If people aren't used to that level of accountability, it can cause friction...} (P35)

Among those other divisions, the Academic Division was identified as of particular interest:

\begin{quote}
    \textit{"
    What we want to do is bring it a little bit more rigor around things like performance of that Academic Plan. Which is getting, not so much push back, just sort of 'hey? We've always done it this way. We've done what we've done.' Well, why do we need this? And that's where it's like, you know, the expectation of the Ministry, the expectation of our Board of Governors, and the expectation of our President is a little bit more towards, you know, evidence-based."} (P35) \label{P35-3}
\end{quote}

But it is not only performance-based data that is absent within the Academic Division. Faculty operate within a complete data silo at the college. When asked about the ways in which data informs learning at the college, non-faculty participants were unable to identify what data faculty use within their roles, with most indicating none at all: \textit{"I'm not particularly sure. I don't think that many do often; they probably look at some of the student data like the grades that they give the students. But I don't think they look at it holistically."} (P33)

Similarly, faculty participants were unaware of the data and data-driven systems used by the college, as well as what decisions are currently supported algorithmically. When asked what data-driven systems are in use, P15, a faculty member who has been with the college full-time since 2014, responded: \textit{"I'd have to guess. I am not a hundred percent sure."} (P15)

Institutional silos are extended to and replicated within the college's data practices. Faculty are kept separate from the college's push towards adopting data-driven processes, even when those processes rely on data coming from the classroom and aim to impact student's academic results, such as in the case of increasing retention. Data-driven systems are being used to automate elements of the student-faculty relationship but are built, implemented, and acted on without faculty knowledge, as discussed in \ref{relationship}.

\subsubsection{Data Processes} \label{Data Processes}
The faculty silo extends to the model design process as well. The EAS, for example, aims to predict a student's academic success - that is, their success in faculty-controlled courses, classrooms and assignments. But when asked if faculty were involved in the design of the model, the the data team-member responsible for its build replied \textit{"for building the model is only me... I only give the scores to the user - the student team..."} (P31) \label{P31-4}

The Student Experience team was similarly hands-off for the model development. P53 explained: \textit{"I mean, it was collaborative, and they listen to us, and entertain our thoughts. But at the end of the day, I trust them to to be able to to do what they need to do. They're the data scientists."}

Once developed, the model output is loaded into the Advisor's CRM Platform. Analysis of the scores themselves and the development of appropriate interventions is outside the scope of the team that built the model and instead falls to the Student Experience team. As the EAS puts students in one of three categories, the Advising team sought to better understand the causes behind an individual student's categorization: was a student flagged as red because their grades were falling or because they began their program in the winter semester? In order to intervene meaningfully, more needs to be understood about a student than just their category or score. To enable advisors to analyze the data at the student or program level, they developed a dashboard to accompany the model's outputs:

\begin{quote}
    \textit{"The data team has created the algorithm. They are crunching the numbers and they are delivering the information to us. And then it's up to our team to actually action it, analyze and see what is this data actually telling us. And how are we going to use it to inform our practice. They might create the score [for the EAS] and those scores may be uploaded into the CRM. But then our job is to help the success advisors understand what that means and then work with them to create the intervention strategy and then to actually follow that through... 
    "} (P53) \label{P53-3}
\end{quote}

While the day-to-day work of a Success Advisor is street-level, working through individual student issues from administrative scheduling issues to connecting them with mental health supports, algorithmic scoring and increased access to data moves their responsibilities towards analyzing the trends within their assigned student populations. A student that has been identified as having a higher probability of failing requires an investigation from the Advisor, who is left to pick apart an opaque and inexplicable output to determine the causes underlying the model's decision and design meaningful interventions for the student.

\subsubsection{Perceptions of Data and Algorithmic Decision-Making} \label{Perceptions of Data and Algorithmic Decision-Making}
Strong discrepancies exist not just in how participants describe the college's challenges in becoming data-driven, but also in how they describe the current state of Centennial College's practices. As discussed \hyperref[Data]{earlier}, many describe data at the college as difficult to access, opaque and inefficiently used. One participant at the Director level described the college's data systems as "\textit{kind of held together with Scotch tape}" (P53). But the team responsible for data at the college views itself as a leader in Ontario's college system: \textit{"I really do feel like in my years here, we've gotten a bit ahead of the curve"} (P35). This disparity might be a question of prioritization. For the the data and research team, algorithmic decision-making is the top priority.

Setting priorities takes on additional importance as the business, academic, and student experience divisions vie for the scarce resources of the data team itself. A Director in Student Experience described how data projects are prioritized: "\textit{At the beginning of each year, the powers that be - committees, working groups - determine what the priorities are going to be. And so if it's on their list, they'll do it, and if it's not on their list they won't do it... 
    } (P53)

One possible driver behind the prioritization of in-house models is the opportunity for commercialization, as described by a member of the executive team: \textit{"If we could get this thing [the International enrollment prediction algorithm] automated, we could commercialize it. That's perfect."}\label{P37-1}

For those in the Business Division of the college, algorithmic decision-making isn't only a tool for efficiency and cost-reduction, but could lead to a possible source of revenue. The emphasis on financial sustainability has made algorithmic and data-driven decision-making a top priority for the college, but for those in the Academic or Student Experience Divisions of the college, the data systems and practices currently in place fall far short of their needs. And as all teams are asked to adopt data-driven practices and algorithmically supported decision making, those with access to data are at an advantage, as we discuss in \ref{weight}.

\section{Discussion} \label{Discussion}
\begin{figure*}[ht]
\caption{The ASP-HEI Cycle - Algorithms, Student Data, and Power in Higher Education Institutions}
\label{fig:ASP-HEI Cycle}
\includegraphics[width=\textwidth]{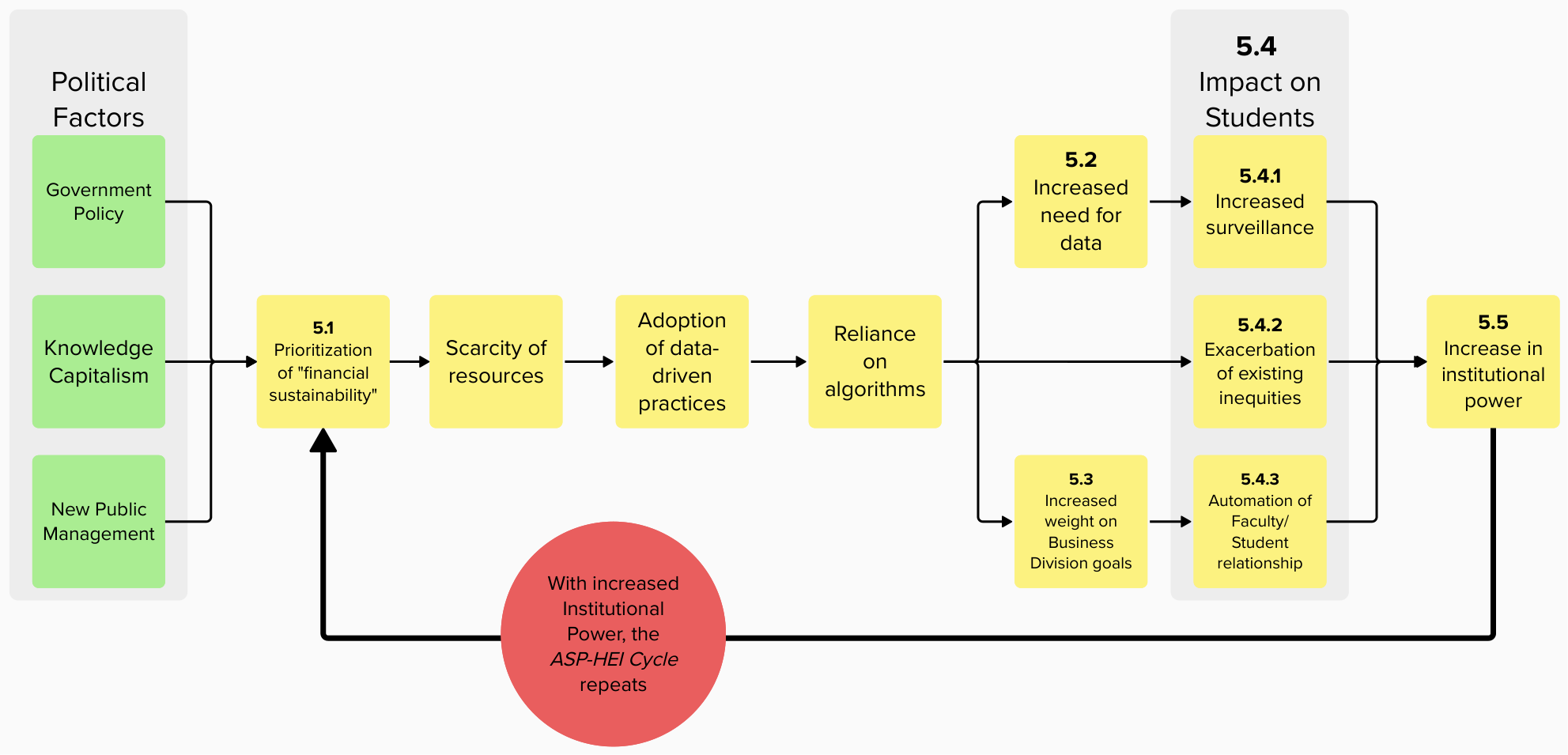}
\end{figure*}
Our results provide insight into the data-driven activities that impact the relationships of stakeholders of a public college. Through our inductive coding, we identified themes (the yellow boxes) that together form  a cycle of increasing institutional power, the \textit{ASP-HEI Cycle} (Fig. \ref{fig:ASP-HEI Cycle}) - Algorithms, Student Data, and Power in Higher Education Institutions. Fig. 1 demonstrates the connections between the themes that emerged, the directional relationship indicated by the arrows.

In this section, we discuss how the prioritization of financial sustainability, resource scarcity, adoption of data-driven practices, along with the reliance on these algorithms as identified in our findings, work to alter the relationships of the college's divisions and ultimately impact students. Each subsection of our discussion aligns with an element on the right side of the ASP-HEI Cycle. Specific quotes from our Results are stated in the format (Participant Number, Quote Number) and are linked directly.

\subsection{Financial Sustainability, Resource Scarcity, and Increased Reliance on Data and Algorithms}\label{financial}
Recent policy by the Government of Ontario \cite{government_of_ontario_published_2023}, as well as a changing educational landscape influenced by \textbf{New Public Management} \cite{lane_new_2000} approaches and \textbf{Knowledge Capitalism} \cite{burton-jones_knowledge_1999}, have pushed HEIs to prioritize financial sustainability. In a report commissioned by the government of Ontario, the blue ribbon panel on the financial sustainability of the postsecondary sector recommends the implementation of a sector-wide accountability framework - formalizing data as a means to achieve financial sustainability across HEIs \cite{harrison_ensuring_2023}. Additionally, the report recommends that "enhance[d] cost efficiency through...  could be realized through increased levels of automation and advancements in digital service delivery" \cite{harrison_ensuring_2023}.
This policy position was reflected at street-level by our participants as an environment of \textbf{resource scarcity} (P20, \aptLtoX[graphic=no,type=html]{4}{\pageref{P20-1}}), with the \textbf{adoption of data-driven practices} highlighted as a proposed remedy (P35, \aptLtoX[graphic=no,type=html]{4}{\pageref{P35-1}}) \textbf{(RQ1)}. This is consistent with prior work in public sector algorithms \cite{saxena_unpacking_2022}. The financial promises of automation through \textbf{algorithmic decision-making} go beyond just the allocation of limited resources, as participants described the college's goal of allocating data-driven savings to new programs and marketing that can generate more revenue (P35, \aptLtoX[graphic=no,type=html]{4}{\pageref{P35-2}}), and even the commercialization of the college's 'trained' algorithmic models (P37, \aptLtoX[graphic=no,type=html]{8}{\pageref{P37-1}}). 
\subsection{Increased Data Needs}The promotion of algorithms as a means to achieve college financial sustainability goals creates an appetite for new data sources. The college's model design and development is an automated process, with models unable to be manually improved or manipulated by the college's data and research team, who are then limited to adding new data as the only way to improve model performance (P31, \aptLtoX[graphic=no,type=html]{6}{\pageref{P31-1}}), resulting in increased surveillance of students and as a result, the loss of self-determination (\ref{surveillance}) \textbf{(RQ2)}.

\subsection{Increased Weight on Business Division Goals}\label{weight} Growing reliance of data-driven practices redistributes power between the college's business, academic, and student experience divisions. As decisions increasingly require data as evidence, those with access to data have an advantage in meeting their priorities. Data-based evidence, however, is often contradictory, as different teams use different data sets, metrics, and definitions to describe the same phenomena, as identified by P23 (P23,\aptLtoX[graphic=no,type=html]{4}{\pageref{P23-1}}) \textbf{(RQ2)}. The data and research team, located within the business division, is primarily responsible for the college's data processes and initiatives. Therefore, the business division has the greatest influence over how the college uses data and what narrative that data support. As an example, the business division identified one goal of putting statistical rigor around the college's Academic Plan (P35, \aptLtoX[graphic=no,type=html]{7}{\pageref{P35-3}}), effectively quantifying the activities of the college's academic division through data controlled by the business division. In this way, institutional power shifts from the academic and student divisions to the business division. There is well-documented risk from over-reliance on algorithms in prior Education research \cite{prinsloo_black_2020, thomas_reliance_2022}; Thomas et al. describe it aptly: "when a measure becomes the target, it ceases to be an effective measure." 

Meanwhile, Coston et al. \cite{coston_validity_2023} point to the issue of achieving validity in designing algorithms for high-stakes decision-making. While the EAS may appear accurate in respect to identifying those students with a higher probability of failing, it may be inaccurate in identifying those students who would most benefit from the college's limited resources \textbf{(RQ1)}.  The academic and student experience divisions have pushed back in large part by decentralizing their data processes from the data and research office and into their own teams (P23, \aptLtoX[graphic=no,type=html]{7}{\pageref{P23-2}})\textbf{(RQ2)}. This creates additional tensions and disparate narratives as siloed teams develop their own metrics and data-practices for measuring the same phenomena, further increasing the challenge of ensuring validity in data-driven and algorithmic systems \cite{coston_validity_2023, jarrahi_principles_2023}.
\subsection{Impact on Students} \label{students}
There is a tangible impact to students from both the increased weight on the business division's goals and the increased need for data at the college. Ultimately, the adoption of data-driven practices and growing reliance on algorithmic decision-making lead to increased student surveillance (as stated by P31, \aptLtoX[graphic=no,type=html]{6}{\pageref{P31-3}}), exacerbation of existing inequities (P31, \aptLtoX[graphic=no,type=html]{6}{\pageref{P31-3}}), and the automation of the faculty/student relationship (P25, \aptLtoX[graphic=no,type=html]{5}{\pageref{P25-1}}), which we discuss in the next sections in detail, and describe how these factors contribute to a cycle of increased institutional power that redefines the relationship between student and school.
\subsubsection{Increased surveillance} \label{surveillance}
The data and research office team's reliance on an auto-ML platform \cite{datarobot_datarobot_2023} provides them with little control of their models. Improving model performance is therefore limited to the addition of new data, as identified by P31 (\aptLtoX[graphic=no,type=html]{6}{\pageref{P31-1}}). That, combined with their established trial-and-error feature selection process, is driving the need for increased student data, and with it, increasing surveillance on students' activities \textbf{(RQ2)}. Although not currently used, participants described future plans to integrate LMS activity into their models (P31, \aptLtoX[graphic=no,type=html]{6}{\pageref{P31-3}}). This LA data is comprised largely of time-stamped logs of student interactions with posted course content: views of videos, frequency of logins, reviews of the syllabus, etc. With more courses and full programs being launched in a fully online, asynchronous format, every interaction a student has with the college could soon be tracked. And with an increase in remote learning \cite{government_of_ontario_ontario_2021} and online proctoring \cite{kimmons_proctoring_2021}, institutional surveillance follows students onto their personal devices and into their homes. The institution's ability to monitor its students is no longer confined to the college campus.
  
Each student is quantified through their admissions data, demographic information, grades, and LA activity, the results of which are used to assign different forms of categorization. Categories are assigned according to the College's undefined and opaque criteria, with no transparency for students \textbf{(RQ1)}. While students are never told the ways in which they are categorized, these categories shape their academic careers and success. In this way, the College's social relations and social power are both organized by its surveillance strategies \cite{lyon_surveillance_2010}. Lyon asserts \cite{lyon_surveillance_2010} that this surveillance can lead to new forms of exclusion, as those surveilled are classified as undesirable. According to Zuboff \cite{zuboff_big_2015}, this can in turn lead to 'anticipatory conformity': students will adjust their learning behaviours so as not to be identified by the college's data-driven and algorithmic practices and risk being classified negatively. While this could include improving study habits and attendance, the opacity of the colleges data systems creates an unknown target for students. As students modify their behaviour in anticipation of how they might be surveilled, institutions are able to control student behaviors to improve the institution's outcomes and goals at the cost of the students self-determination and autonomy \cite{brown_whose_2020}. This aligns with prior Education research \cite{prinsloo_black_2020} and that of the broader research community. Zuboff \cite{zuboff_big_2015}, for example, describes a surveillance society in which "human autonomy is irrelevant and the lived experience of psychological self-determination is a cruel illusion." The erosion of self-determination, as described by Zuboff, is fundamentally at odds with the college's claim of "valu[ing] the diverse profiles of our learners." \cite{centennial_college_vision_2023} As it strives towards inclusive classrooms that recognize students' individuality \cite{centennial_college_vision_2023}, the college's own data practices could lead students to attempt to conform for fear of being labelled undesirable.
 \subsubsection{Exacerbation of existing inequities} \label{inequity}
The impacts of data-driven practices, algorithmic decision-making, and surveillance are not felt equally across the student body. Biases in training data, discrimination in the model's output, overemphasis on quantitative metrics and lack of context, and lack of transparency all exacerbate existing inequities in the College \textbf{(RQ1)}. Participants from the data and research office team indicated that there was no process in place for checking the training data and the model's outputs for bias (P31, \aptLtoX[graphic=no,type=html]{6}{\pageref{P31-3}}), while the Student Experience team that implements the model's decisions was confident those processes were in place (P53, \aptLtoX[graphic=no,type=html]{6}{\pageref{P53-1}}). Participants from both the academic and student experience divisions described the necessity of context for any decision involving students. However, for Advisors, the EAS provides a low-context output of three possible categories: red, yellow, or green (P25, \aptLtoX[graphic=no,type=html]{5}{\pageref{P25-1}}). Students are unable to contest the model's categorization of probability of academic achievement because they are unaware of its existence. And should an algorithmically-supported decision at the college cause harm, there is likely no clear recourse for students as data ownership is unclear (P53, \aptLtoX[graphic=no,type=html]{6}{\pageref{P53-2}}) and the model's decisions are opaque and unexplained (P53, \aptLtoX[graphic=no,type=html]{8}{\pageref{P53-3}}). No one can explain to a student how the model made a decision or who is responsible for that decision, and that is assuming the student is able to determine an algorithm was used at all \textbf{(RQ2)}.
    
The EAS is trained on historical college data, with the aim of predicting whether or not the student will successfully complete their program (P25, \aptLtoX[graphic=no,type=html]{5}{\pageref{P25-1}) \textbf{(RQ1)}}. Any bias and inequities that existed within the colleges retention history, and therefore its training data, will be formalized within the model as the model replicates past outcomes. And the model's decisions once implemented, in this case, determining who is given access to support and interventions, effectively becomes policy. Biases formalized within the algorithmic decision-making process become manifestations of the college's institutional power \cite{alkhatib_street-level_2019}.

Academic persistence is complex and highly contextual, and students from underrepresented and equity-seeking groups are more likely to leave higher education without graduating \cite{finnie_patterns_2012}. Without careful and ongoing review of the training data and model output, systemic inequities felt by underrepresented student groups could be perpetuated and heightened by reliance on algorithmic decision-making \cite{suresh_framework_2021,mehrabi_survey_2021}. As the model learns from the college's history, it learns that disabled, low-income, or indigenous students \cite{finnie_patterns_2012}, for example, are less likely to be successful without an understanding of the wider systemic context. Prior research indicates that when algorithmically predicting student success, disadvantaged student sub-populations are disproportionately discriminated against \cite{yu_towards_2020}. Brown et al. \cite{brown_whose_2020} describe this as digital redlining, education policies, investments, and technology practices working in concert to discriminate against certain groups and maintain class boundaries. For example, bias in the data could cause higher Type II errors (false negatives) for certain groups. As the model output determines a student's access to interventions, this would leave some students out of student support initiatives based on their race, disability status, or income. The support initiatives include opportunities to be connected to free counselling services and academic supports including personalized outreach, all focused on helping students graduate (P25). The potential impact of algorithmic discrimination could follow a student long after the decision is made and affect the core activities of their college career: their grades, ability to graduate, and future career trajectories.

Students who are part of equity-seeking groups are also more greatly impacted by surveillance. International students in Canada, for example, are surveilled by their institutions as part of the Federal Government's ‘international student compliance regime’, a tool that sorts, identifies and governs the most desirable students and in doing so, reinforces the college's social and institutional power \cite{brunner_higher_2023}. As educational surveillance grows in the pursuit of training data, there is no guarantee to students that that data won't be used for other purposes downstream. \cite{brown_whose_2020} For example, while Centennial College uses student data with the goal of improving retention, at at least one HEI, it was used to stop those identified students from enrolling at all. Mount St. Mary’s University President Simon Newman proposed using predictive student analytics to weed out students unlikely to graduate, as a way  to "drown the bunnies... put a Glock to their heads” \cite{johnson_structural_2018}. While especially egregious, it highlights how data can be used in ways beyond those identified when it was collected to limit students' agency.

And as the majority of colleges in Ontario are publicly-assisted and deeply controlled by the provincial government \cite{warren_limits_2020}, those downstream purposes could include monitoring compliance to other government regulations, as seen with the international student compliance regime.
\subsubsection{Automation of Faculty/Student relationship} \label{relationship}
\begin{figure}[ht]
\caption{Screen capture of output categories of the EAS model.}
\label{fig:categories}
\includegraphics[width=\columnwidth]{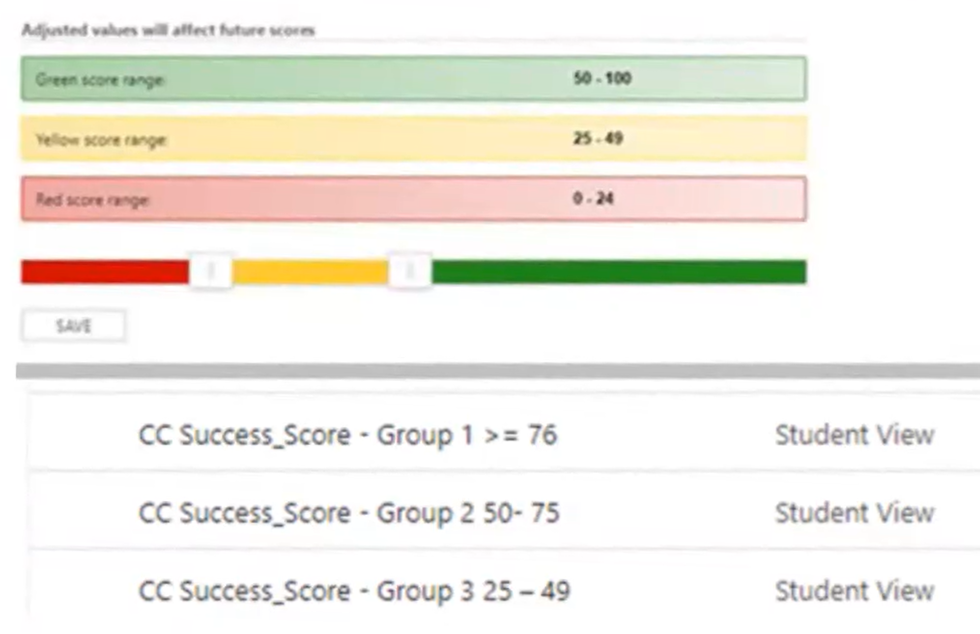}
\end{figure}

The faculty-student relationship plays a key role in promoting student success and increasing retention \cite{lillis_faculty_2011}. The more interactions that faculty have with students, the more likely the student is to remain enrolled in an institution \cite{tinto_misconceptions_1989}. The stated promise of the EAS is to automate that process, replacing a faculty and student-driven process with an algorithmic output \textbf{(RQ1, RQ2)}. As per our findings, faculty are not involved in any step of this process beyond providing the grades that are used in some iterations of the model as training data (P31, \aptLtoX[graphic=no,type=html]{7}{\pageref{P31-4}}). This automation is designed to allow the college to target interventions, track students' scores as they progress through their academic careers and the corresponding algorithms, and ultimately, reduce attrition and increase revenue. But by automating the identification of at-risk students, the college has formalized the meaning of "at-risk" without ever defining "at-risk". The most recent implementation of the model's output declares 'at-risk' as the bottom quartile of probabilities of graduating (Fig. \ref{fig:categories}). 
For incoming students, the probability is mostly determined by the interaction of the model's three most important features: the student's age, program, and semester of enrollment (Fall, Winter, or Spring/Summer) (as described by P25, \aptLtoX[graphic=no,type=html]{6}{\pageref{P25-2}}). By implementing the model, the college has formalized the risk factors of success for new students as elements of registration and timing. There have been no attempts by the college's data team to understand the underlying relationship, and further, program offerings and intake semesters are controlled by the college. Despite this, the model's decisions trigger only student-level interventions. The model is fundamentally a deficit-based system, focused on quantifying risk but not on improving students' lives \cite{saxena_framework_2021}.

Traditionally, the act of flagging struggling students and intervening in their learning fell to faculty who relied on the social capital of a relationship built on casual and formal interactions \cite{tinto_misconceptions_1989}, interactions that play a crucial role in retention \cite{lillis_faculty_2011}. Once automated by algorithmic decision-making, faculty no longer play a role and what was once \textit{a relationship with} students becomes \textit{a process done to} students. Students lose agency during a decision-making process critical to their success. For faculty, what was once a process initiated by the academic division becomes one controlled by the business division. By removing faculty and students from this process, the college risks greater harms to students and lowering trust in the system itself \cite{zhang_deliberating_2023}. And further, teaching and learning are redefined as business-driven activities. As the business division measures, models on, and reports on those teaching and learning activities, the goal of the classroom becomes lowering the probability of a student failing as determined by an algorithm, not meeting learning outcomes or helping students' reach personal goals, and certainly not "serving the needs of society through the development of knowledge as a public good." \cite{peters_higher_2007}

\subsection{Increase in Institutional Power} \label{power}
As discussed in section \ref{financial} above, in an attempt to allocate resources and increase revenue (P35, \aptLtoX[graphic=no,type=html]{4}{\pageref{P35-2}}), academic and student experience activities and processes at the college are shifted to the business division (\ref{weight}), and important elements of a student's academic and college experience are formalized within opaque black boxes (P53, \aptLtoX[graphic=no,type=html]{8}{\pageref{P53-3}}) \textbf{(RQ1, RQ2)}. The college's data-driven processes and algorithms effectively replicate the power structures that already exist in the college as institutional power is amalgamated and centralized through increased surveillance (\ref{surveillance}), exacerbation of existing inequalities (\ref{inequity}), and the automation of the faculty/student relationship (\ref{relationship}), the impact of which is felt disproportionately by the college's most vulnerable students. With increased power and persistent external austerity measures exerted on it, the college is further empowered to seek even greater cost savings and increased revenue, beginning the cycle again (Fig. \ref{fig:ASP-HEI Cycle}). This aligns closely with prior Educational research by Piattoeva et al. who theorize that in HEIs, data acts as "a silent, implicit authority in which the exercise of power is difficult to discern" \cite{piattoeva_escaping_2020}.

The data practices and algorithms of higher education rely on the strategies, policies, and technologies that enable increased surveillance, measurement, evaluation, and governance \cite{brown_whose_2020}. As evidenced by the works of Benjamin \cite{benjamin_race_2019}, Cohen \cite{cohen_biopolitical_2018}, and Zuboff \cite{zuboff_big_2015}, these elements ultimately work together to reinforce institutional practices that extract from students and build resources towards the internal and external goals of efficiency and commercialization.

\subsection{Implications for Algorithmic Decision-Making in Public Higher Education}

The ADMAPS framework \cite{saxena_framework_2021} was used deductively to thematically code our data. It situates our results within prior SIGCHI work within the public sector and provides insight into the broader implications of our findings for algorithm design in higher education. 

\paragraph{Balancing Strength-based and Deficit-based Approaches}
As HEIs move towards algorithmic decision-making, those algorithms largely constitute risk-assessment models \cite{mcconvey_human-centered_2023}. These models aim to mitigate risk to the institution and by doing so, attempt to quantify core educational activities and risk losing important context. By removing Advisor and faculty discretion from the decision-making process, institutions risk increased potential for harm and a loss in trust in data-driven practices across stakeholders. By combining risk assessment with strength-based algorithms as proposed by Saxena et al. \cite{saxena_framework_2021}, institutions could center the student's experience and success while balancing the need for financial sustainability.

\paragraph{Holistic Assessments, Not Deterministic Scoring Lead to Improved Decisions}
Deterministic scoring such as in the case of the EAS effectively boils students' potential down to a single value. As identified in prior research \cite{saxena_framework_2021}, this has the potential to incentivize 'anticipatory conformity' \cite{zuboff_big_2015}, leading students to modify their educational behaviours to avoid being classified negatively, resulting in a loss of agency for the student and risking discrimination of student groups. Anticipatory conformity is at odds with the colleges commitments to flexible and inclusive learning \cite{centennial_college_universal_2020}. By assessing students' needs holistically, with an understanding of the underlying lived experiences of individual students, algorithmic decision supports can be "developed that support and/or streamline bureaucratic processes and augment the quality of human discretionary work." \cite{saxena_framework_2021}

\section{Limitations and Future Work}
This paper provides an in-depth ethnographic case study of the data-driven and algorithmic decision-making practices of a public college and contributes a model for conceptualizing the relationship between external political factor and data practices, and their impact on students and institutional power. There are several limitations that provide opportunities for future research to expand upon this work. Firstly, in this paper we discuss the potential for bias in the data, models and implementation of algorithmic decision-making but we do not deconstruct or analyze the those elements directly. Secondly, the study was limited to discussion of algorithms that are designed and trained directly by the college's data team. There may be algorithmic tools in use that are built into third-party platforms purchased and used by the college. Further research should consider the impact of those off-the-shelf predictive tools and their use in HEIs. Therefore, researchers should focus on analysing evaluating the datasets, models and implementations used in practice in higher education settings such as our research site, and the role of third-party predictive systems within the \textit{ASP-HEI Cycle}. Additionally,  our findings may not be generalizable to other HEIs; future research is required to test the applicability of the ASP-HEI cycle beyond our research site.

\section{Conclusion}
We conducted an in-depth ethnographic case study of data and algorithms in use at a public college in Ontario, Canada. Through our qualitative analysis, we: a) identified the data, algorithms, and outcomes in use at the college, b) assessed how the college's processes and relationships support those outcomes, and c) highlighted the different stakeholders' perceptions of the college's data-driven systems. We adopted the ADMAPS framework as a theoretical lens through which to code our qualitative data and aligned our findings with the ADMAPS dimensions of human discretion, algorithmic decision-making, and bureaucratic processes. Our research revealed that the increasing reliance on algorithmic decision-making has led to heightened surveillance of students, the exacerbation of preexisting inequities, and the automation of aspects of the relationship between faculty and students. Ultimately, our study uncovered a recurring pattern of heightened institutional power perpetuated by algorithmic decision-making and driven by the prioritization of financial sustainability.
\begin{acks}
Opinions, findings, and conclusions in expressed in this paper are those of the authors and do not necessarily reflect the views of our research site partner. We would like to thank our collaborators and study participants at Centennial College for allowing us to conduct this ethnography. Additionally, we thank the anonymous reviewers whose suggestions and comments helped improve this manuscript. This research was supported by the Natural Sciences and Engineering Research Council of Canada's Early Career Discovery Grant.
\end{acks}

\bibliographystyle{ACM-Reference-Format}
\bibliography{main}

\appendix
\end{document}